\documentclass[PRE]{revtex4}

\usepackage{graphicx}

\usepackage{amssymb}

\usepackage[figuresright]{rotating}

\newcommand{\be}{\begin{equation}}
\newcommand{\ee}{\end{equation}}

\begin{document}

\title{Phase separation on a hyperbolic lattice}


\author{Jesse  Raffield$^1$,
Howard L.\ Richards$^2$,
James Molchanoff$^2$,
Per Arne Rikvold$^1$}

\affiliation{$^1$ Department of Physics, 
Florida State University, Tallahassee, FL 32306-4350, U.S.A.\\
$^2$  Department of Physics and Physical Science, 
Marshall University, Huntington, WV 25755, U.S.A.}

\begin{abstract}
We report a preliminary numerical study by kinetic Monte Carlo simulation 
of the dynamics of phase separation following a quench from high 
to low temperature in a system with a single, conserved, 
scalar order parameter (a kinetic Ising ferromagnet) confined to a 
hyperbolic lattice. The results are compared with simulations of the same 
system on two different, Euclidean lattices, in which cases we observe 
power-law domain growth with an exponent near the theoretically known value 
of 1/3. 
For the hyperbolic lattice we observe much slower domain growth, 
consistent to within our current accuracy 
with power-law growth with a much smaller exponent near 0.13. 
The paper also includes a brief introduction to non-Euclidean lattices 
and their mapping to the Euclidean plane. 
\end{abstract}







\maketitle 

\noindent
{\bf Keywords:} Hyperbolic lattice;
Phase separation;
Coarsening;
Kinetic Monte Carlo simulation.

\section{Introduction}
\label{sec-Int}

The geometry most familiar to condensed-matter physicists is the Euclidean one with its vanishing Gaussian curvature 
\cite{Note1}. The circumference and area of a circle 
of radius $\rho$ in the Euclidean plane are the well-known power laws, 
$C(\rho) = 2 \pi \rho $ and $A( \rho ) = \int_0^\rho C(r)dr = \pi \rho^2$, 
respectively. Almost equally familiar are elliptic or spherical surfaces, exemplified in the macroscopic world by the Earth's 
surface and in the nanoscopic world by carbon buckyballs. 
These closed surfaces have positive Gaussian curvature, $\kappa > 0$.
Hereafter using dimensionless units such that $| \kappa | = 1$,  
the circular circumference and area are analogously given by 
$C(\rho ) = 2 \pi \sin \rho $ and $A(\rho ) = 2 \pi (1 - \cos \rho)$. 

More exotic to most is probably the hyperbolic geometry with its negative 
Gaussian curvature, $\kappa < 0$. In this case the dimensionless circular 
circumference and area are exponentially divergent:
$C(\rho ) = 2 \pi \sinh \rho$ and $A(\rho ) = 2 \pi (\cosh \rho - 1)$. 
The best known example is the Minkowski metric of relativistic spacetime. However, hyperbolic surfaces have recently 
been studied in nanoscience as well, including junctions of several carbon nanotubes \cite{PARK03} and anisotropic lipid 
membranes \cite{GIOM12}. Percolation on hyperbolic lattices has also been studied \cite{GU12}. 

In this paper we present preliminary results on a 
comparison of the dynamics of pattern 
formation during phase separation (spinodal decomposition) in 
media confined to Euclidean and hyperbolic surfaces. 
As our example we use an $S=1/2$ ferromagnetic Ising model on a 
regular lattice embedded in the surface, and we study the time 
evolution of the characteristic pattern length following a quench from 
infinite temperature to one well below the model's critical temperature. 

The rest of the paper is organized as follows. 
In Sec.~\ref{sec-Map} we describe the Poincar{\'e} disk mapping used to 
map patterns on a hyperbolic surface onto a Euclidean plane. Next, in 
Sec.~\ref{sec-Hyp}, we describe the lattices generated by regular 
tesselations of Euclidean, spherical, and hyperbolic surfaces. 
The phenomenology
of phase separation is briefly reviewed in Sec.~\ref{sec-Sep}, and the 
methods of simulation and data analysis are discussed in Sec.~\ref{sec-Sim}. 
Our numerical results are presented in Sec.~\ref{sec-Num}, and a final 
discussion and suggestions for future work are given in Sec.~\ref{sec-Disc}.

\section{Poincar{\'e} disk mapping of non-Euclidean surfaces}
\label{sec-Map}

\begin{figure}[t] 
\vspace{-0.3truecm}
\begin{center}
\includegraphics[angle=0,width=.9\textwidth]{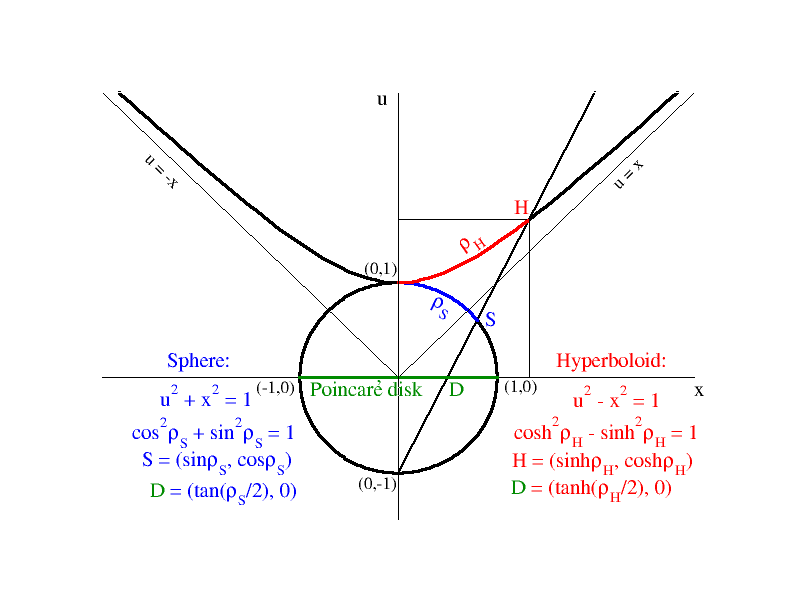} 
\end{center}
\vspace{-2.0truecm}
\caption[]{
The Poincar{\'e} disk mappings from a unit hyperboloid of revolution 
($\kappa = -1$) and a unit sphere ($\kappa = +1$) to a Euclidean plane. 
The $y$ coordinate points into the page and has been suppressed in our 
notation, describing 
the position of a point as $(x,u)$. See discussion in Sec.~\ref{sec-Map}.  
}
\label{fig-Poin}
\end{figure}

Patterns on a non-Euclidean surface cannot be reproduced on a Euclidean surface without distortion; angular, linear, 
or both. No single rendering is ideal in every respect, as evidenced by the many different geographic map 
projections developed over the centuries. The projection of the hyperbolic plane that we use in this paper is known as 
the Poincar{\'e} disk mapping and is illustrated in Fig.~\ref{fig-Poin}. 
It consists of a Euclidean plane, a unit sphere ($ \kappa = +1$), and a 
unit hyperboloid of revolution ($ \kappa = -1$) \cite{CRIA01}. 
The plane forms the equatorial plane of the 
sphere, and the hyperboloid rests with its 
apex on the North Pole of the sphere, (0,1). 
A straight line through the South Pole, (0,$-1$), connects the points 
H on the hyperboloid and
S on the sphere with their mapping D on the plane. The region of the equatorial plane inside the sphere is the 
Poincar{\'e} disk. It is easily seen from Fig.~\ref{fig-Poin} that 
an arbitrary point H on the hyperboloid is mapped 
{\it inside} the Poincar{\'e} disk, leading in the map to an exponential 
contraction of lengths far away from the apex of the hyperbola. 
The mapping from H to D is conformal (angle-preserving), and geodesics on the hyperboloid are mapped onto circles 
on the disk that meet its edge at straight angles. 
A little calculus shows that the ``hyperbolic radius" of H, 
$\rho_{\rm H}$, is the arc length of the geodesic from 
(0,1) to H, calculated with the Minkowski metric, $ds = \sqrt{dx^2 - du^2}$. 
(Analogously, the ``spherical radius" of S, $\rho_{\rm S}$, is the arc length of the geodesic from (0,1) 
to S, calculated with the Euclidean metric, $ds = \sqrt{dx^2 + du^2}$.) 
Other equations relevant to the mappings are included in Fig.~\ref{fig-Poin}.

\section{Euclidean, elliptic, and hyperbolic lattices}
\label{sec-Hyp}

\begin{figure}[t] 
\begin{center}
\includegraphics[angle=0,width=.9\textwidth]{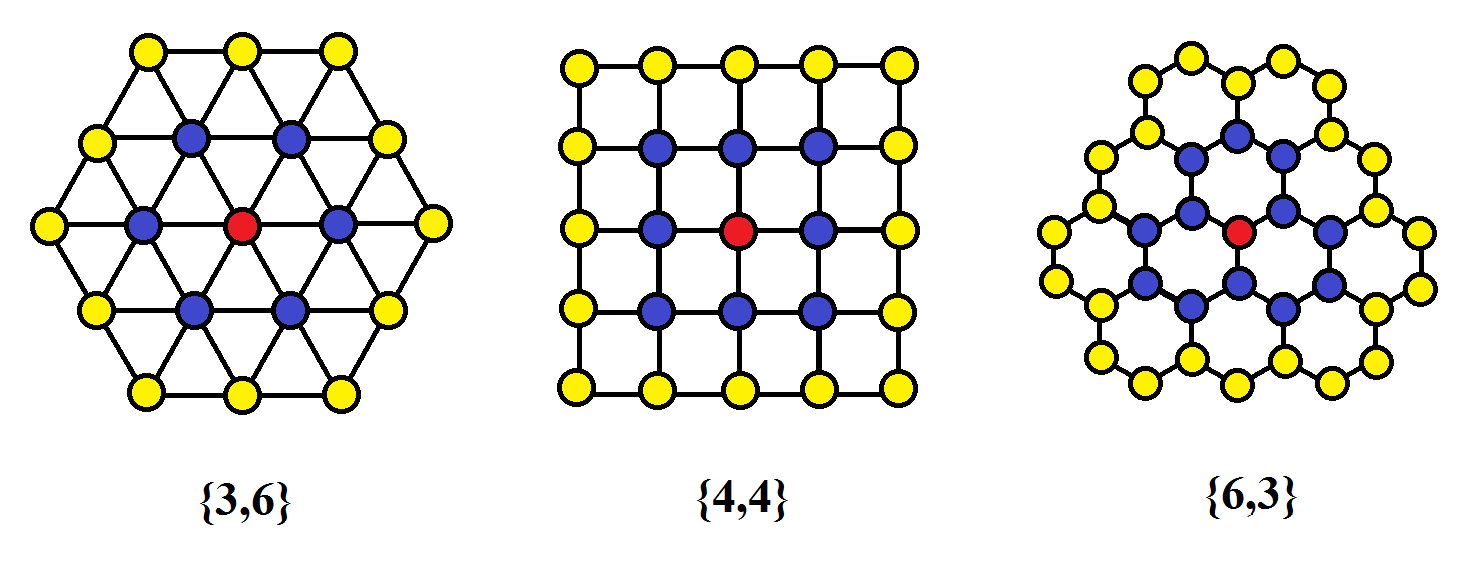} 
\end{center}
\vspace{-0.8truecm}
\caption[]{
The three regular Euclidean lattices $\{p,q\}$ of $q$ $p$-gons meeting at each lattice site. Here, each is 
shown with $r=2$ layers of polygons: $\{3,6,2\}$, $\{4,4,2\}$,
and $\{6,3,2\}$. The $\{3,6\}$ and  $\{6,3\}$ lattices are each other's duals, while  $\{4,4\}$ is self-dual. 
}
\label{fig-EuTess}
\end{figure}

\begin{figure}[h] 
\begin{center}
\includegraphics[angle=0,width=.4\textwidth]{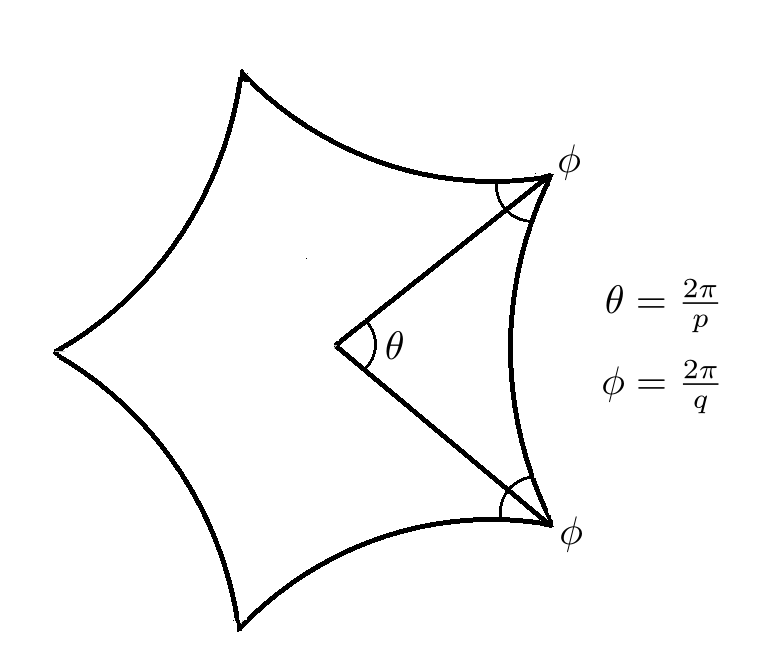} 
\end{center}
\vspace{-0.8truecm}
\caption[]{
Splitting a hyperbolic pentagon into five isosceles triangles. 
}
\label{fig-decomp}
\end{figure}

To construct lattices embedded in Euclidean and non-Euclidean surfaces, we consider regular tessellations of such 
surfaces by regular polygons. A lattice created by this procedure, such that $q$ regular 
$p$-gons meet at every lattice site is characterized by its Schl{\"a}ffli symbol, $\{p,q\}$ \cite{GU12}. A lattice of finite 
size is often denoted by the amended Schl{\"a}ffli symbol, $\{p,q,R\}$, where $R$ is the number of concentric layers of 
$p$-gons surrounding the central site. The only three regular tessellations of the Euclidean plane are shown in 
Fig.~\ref{fig-EuTess}. 
Any regular $p$-gon can be decomposed into $p$ isosceles triangles that meet 
at its centroid. This is not only true for Euclidean lattices. 
An illustration for the hyperbolic case is shown in Fig.~\ref{fig-decomp}. 
Each triangle has apex angle $\theta$ and basal angles $\phi/2$.

For the Euclidean plane, 
the interior angles of a triangle must always sum to $\pi$. Thus, 
\be
2\pi / p + 2\pi / q = \pi \Leftrightarrow (p-2)(q-2) = 4 \;.
\label{eq-Eu}
\ee
This proves that the only combinations of 
integer $p$ and $q$ compatible with Euclidean geometry are 
$\{3,6\}$, $\{4,4\}$, and $\{6,3\}$.

Similarly, for the elliptic plane, 
\be
2\pi / p + 2\pi / q > \pi \Leftrightarrow (p-2)(q-2) < 4 \;.
\label{eq-El}
\ee
Again it is clear that the number of possible regular tessellations is finite. 
In fact, they correspond to the five Platonic solids, $\{3,3\}$, $\{3,4\}$, $\{3,5\}$, $\{4,3\}$, and 
$\{5,3\}$ \cite{COXE63}. 

\begin{figure}[t] 
\begin{center}
\includegraphics[angle=0,width=.9\textwidth]{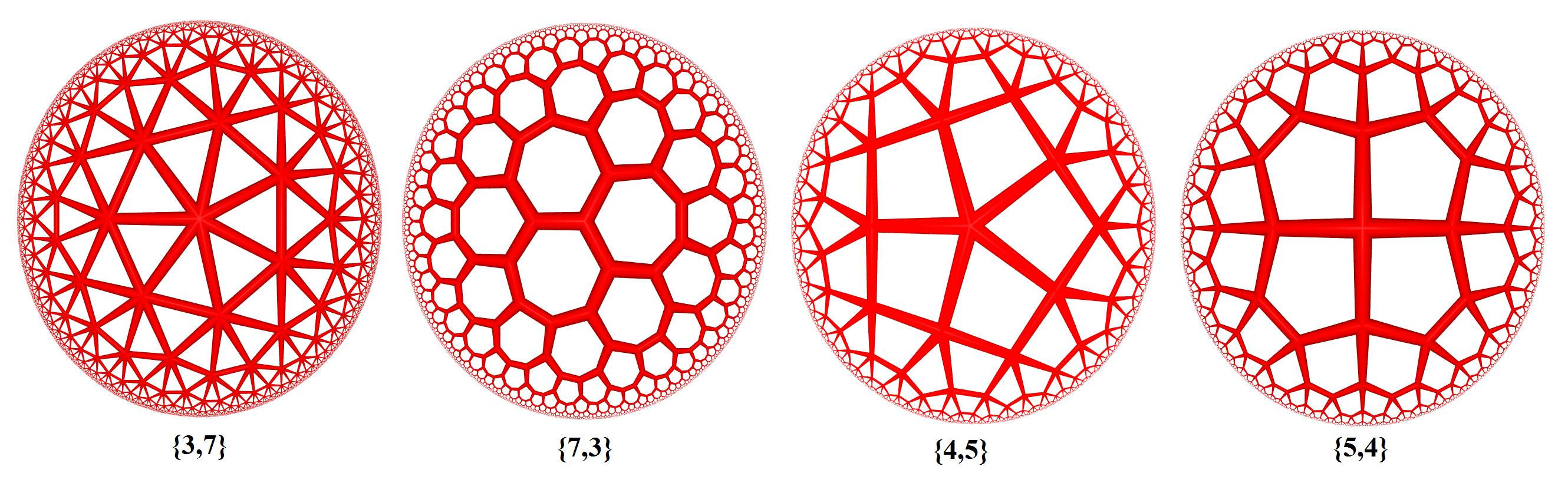} 
\end{center}
\vspace{-0.8truecm}
\caption[]{
Some examples of the infinite number of hyperbolic lattices, projected onto the Poincar{\'e} disk. 
The exponential length contraction in the projected image of large $\rho_{\rm H}$ is clearly evident.
As for the Euclidean and elliptic geometries, $\{p,q\}$ and $\{q,p\}$ are 
duals.
}
\label{fig-H}
\end{figure}

\begin{figure}[h] 
\begin{center}
\includegraphics[angle=0,width=.6\textwidth]{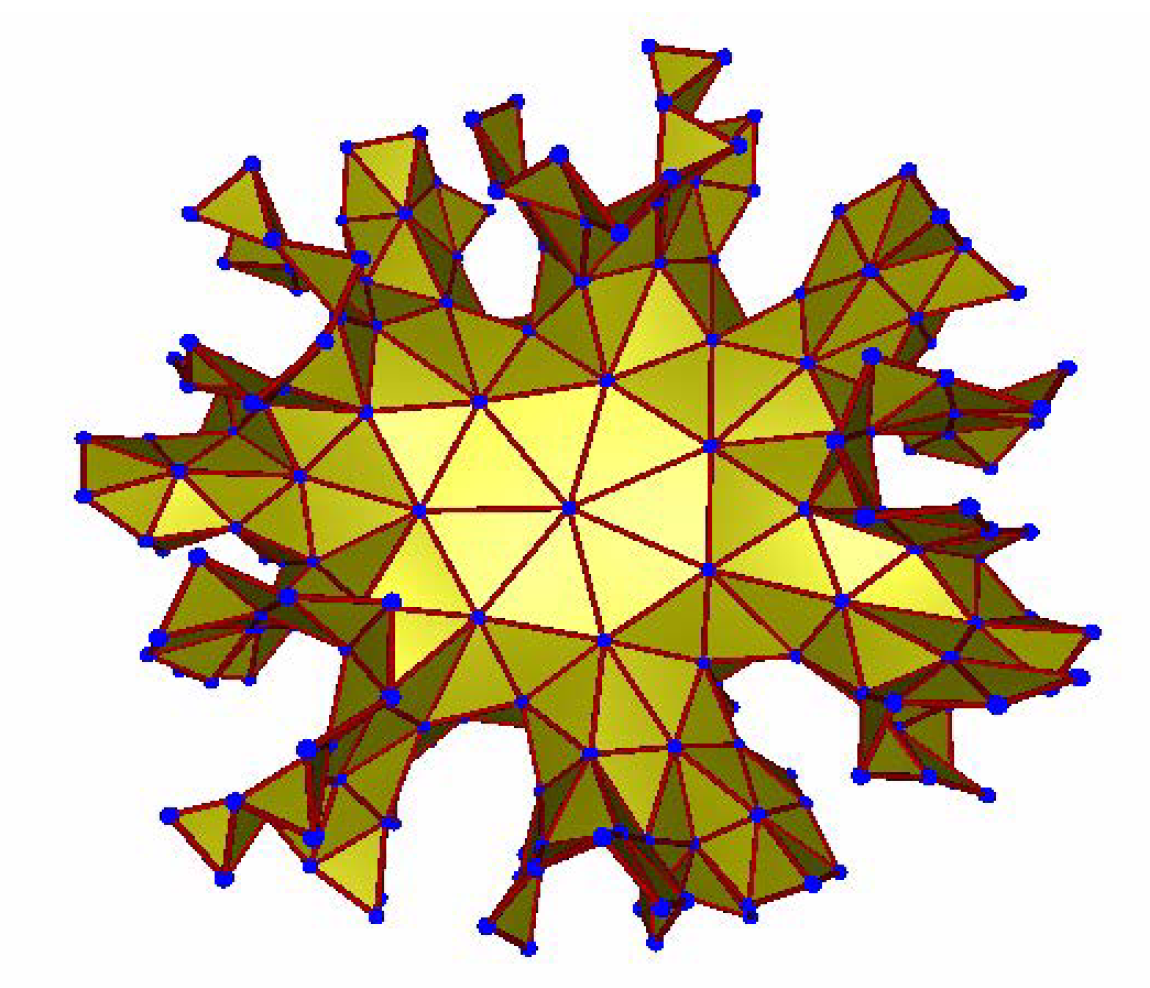} 
\end{center}
\vspace{-0.8truecm}
\caption[]{
Perspective image of a small 
$\{3,7,4\}$ lattice embedded in a three-dimensional 
Euclidean space. The exponential divergence of the hyperbolic circumference 
with radius prohibits such embedding without overlaps for large lattices. 
}
\label{fig-373D}
\end{figure}

For the hyperbolic plane, on the other hand, 
\be
2\pi / p + 2\pi / q < \pi \Leftrightarrow (p-2)(q-2) > 4 \;.
\label{eq-H}
\ee
Consequently, the number of possible regular tessellations is infinite. 
Some examples are shown in Figs.~\ref{fig-H} and \ref{fig-373D}. 
As the size of the lattice increases, embedding without overlaps 
into a three-dimensional Euclidean space becomes impossible. Fascinating 
images of models of hyperbolic planes created by crocheting can be found in 
Ref.~\cite{TAIM09}.  

\section{Phase separation}
\label{sec-Sep}

Phase separation occurs when a binary mixture is quenched from a high temperature into the phase-coexistence region 
below its critical temperature. As coherent regions of the two coexisting 
phases form and grow after the quench, the length 
scale characterizing the typical domain size increases algebraically with 
time as 
\be
\xi \sim t^n\;.
\label{eq-xi}
\ee
The growth exponent 
$n$ depends on the symmetries and conservation laws governing the 
dynamics. For the case of two-phase 
coexistence and a constant ratio of the volumes of the two phases, $n = 1/3$. 
This situation is known in the 
terminology of critical dynamics as  Lifshitz-Slyozov dynamics \cite{LIFS62} 
or Model B \cite{HOHE77}. However, these results 
implicitly assume a Euclidean geometry, and we are not aware that the 
prediction has yet been tested in the hyperbolic case. 
A numerical test is the purpose of the work presented here.

\section{Simulation and data analysis}
\label{sec-Sim}

We consider the phase separation that occurs when a $S=1/2$ Ising 
ferromagnet with spins $s_i = \pm 1$ 
placed at the vertices of the $\{p,q\}$ lattice is 
quenched from a high temperature to one well below its critical temperature. 
The Hamiltonian is given by 
\be
{\mathcal H} = - J \sum_{\langle i,j \rangle} s_i s_j \;.
\label{eq-Ha}
\ee
Here, $J > 0$ is the ferromagnetic interaction constant, and the 
sum runs over all nearest-neighbor pairs. The coordination number in a 
$\{p,q\}$ lattice is $q$. We will use units such that Boltzmann's constant 
and $J$ both equal unity.

While the phase transition at the critical temperature $T_c$ for 
this model on Euclidean lattices belongs to the two-dimensional Ising 
universality class, on hyperbolic lattices it belongs to the mean-field universality class 
\cite{KRCM08,GEND12}. In either case, the value of $T_c$ increases with $q$. 
($T_c = 2/\ln(1+\sqrt{2}) \approx 2.269$ for $\{4,4\}$, 
$4/\ln 3 \approx 3.641$ for $\{3,6\}$, and $\approx 5.5$ for $\{3,7\}$ \cite{GEND12}.)
To minimize surface effects, we use periodic boundary conditions for the 
two Euclidean lattices. Unfortunately we are not aware of a method for 
doing so in the hyperbolic case, and consequently we simulate 
the $\{3,7\}$ lattice with free boundary conditions. 

\begin{figure}[t]
\begin{center}
\includegraphics[angle=0,width=.48\textwidth]{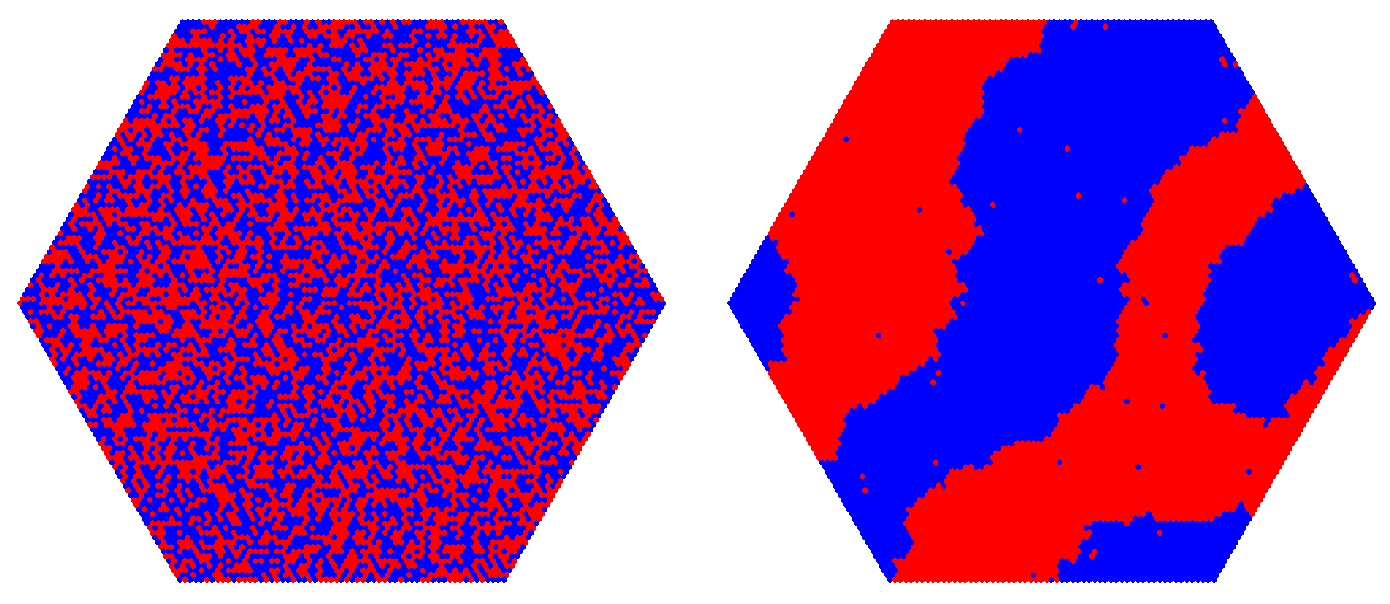}
\includegraphics[angle=0,width=.48\textwidth]{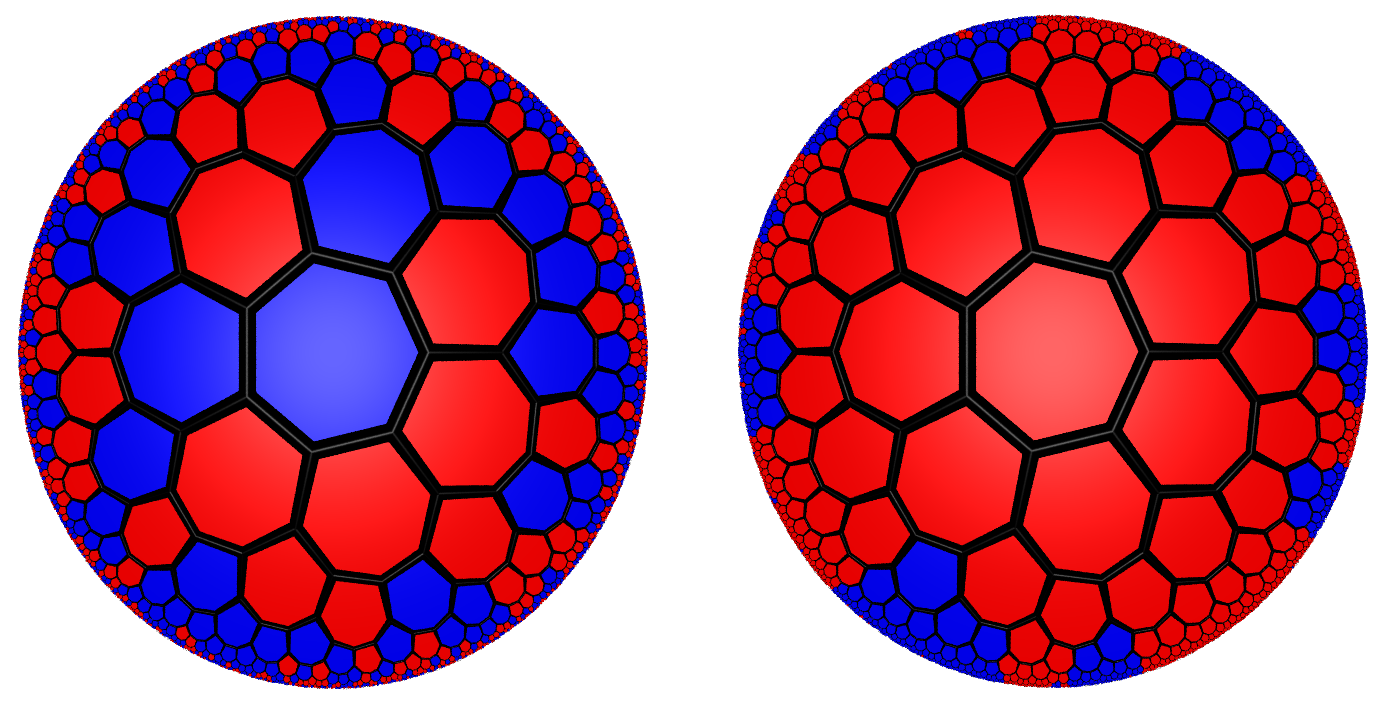}
\end{center}
\caption[]{
Snapshots of the $\{3,6,60\}$ (left) and $\{3,7,8\}$ lattices (right) 
in their initial, disordered states and at times near the end of the 
simulation runs. $T = 2.0$, in both cases well below $T_c$. 
Small thermal fluctuations in the bulk phases are noticeable for the $\{3,6\}$ case. 
}
\label{fig-snaps}
\end{figure}

The initial state is a random distribution of up and down spins, subject only 
to the constraint of a vanishing order parameter, $\sum_i s_i = 0$. 
The time evolution is obtained from a kinetic Monte Carlo (MC) simulation
by the order-parameter conserving Kawasaki dynamics \cite{KAWA72}. 
This algorithm consists in randomly choosing a nearest-neighbor 
spin pair and checking if the two spins are different. If they are equal, 
a different pair is chosen. If the spins are different, they are exchanged 
with the Metropolis probability, 
\be
P_{\rm ex}(s_i,s_j) = \min[1,\exp(- \Delta E / T)] \;,
\label{eq-Kawa}
\ee
where $\Delta E$ is the energy change that would result from a successful 
spin exchange. In a system consisting of $N$ spins, $N$ random choices of a 
spin pair constitute the MC time unit, one MC step per spin (one MCSS).   
Snapshots of the $\{3,6\}$ and $\{3,7\}$ lattices in their initial, disordered 
states and at $t = 10^6$ MCSS, 
when macroscopic bulk phase domains are well developed, 
are shown in Fig.~\ref{fig-snaps}. 

\begin{figure}[ht]
\vspace{-0.4truecm}
\begin{center}
\includegraphics[angle=0,width=.6\textwidth]{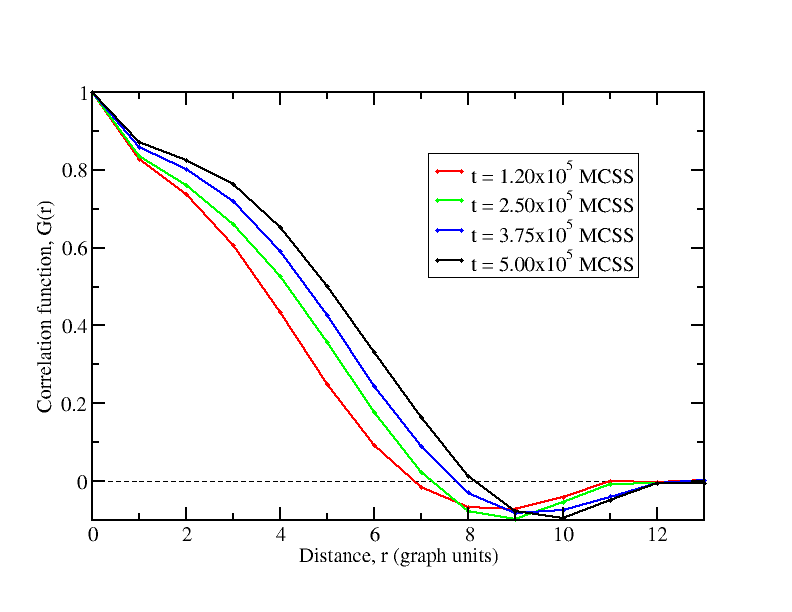}
\end{center}
\vspace{-1.0truecm}
\caption[]{
The spin correlation function $G(r)$ for a $\{3,7,7\}$ 
lattice at different times.
The correlation length $\xi(t)$ is estimated as the distance corresponding to the first zero crossing 
of $G(r)$, as discussed in the text. 
}
\label{fig-corrf}
\end{figure}
\begin{figure}[t]
\vspace{-0.4truecm}
\begin{center}
\includegraphics[angle=0,width=.57\textwidth]{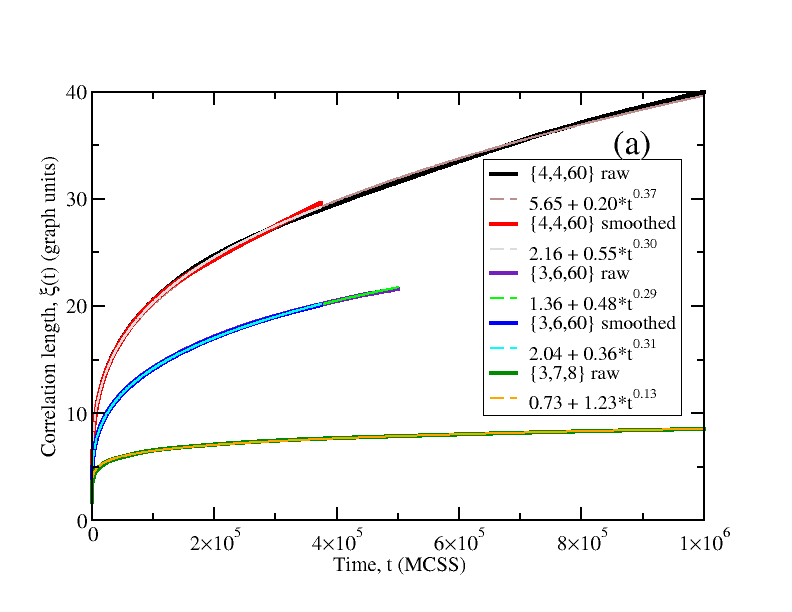} \\
\vspace{-0.5truecm}
\includegraphics[angle=0,width=.57\textwidth]{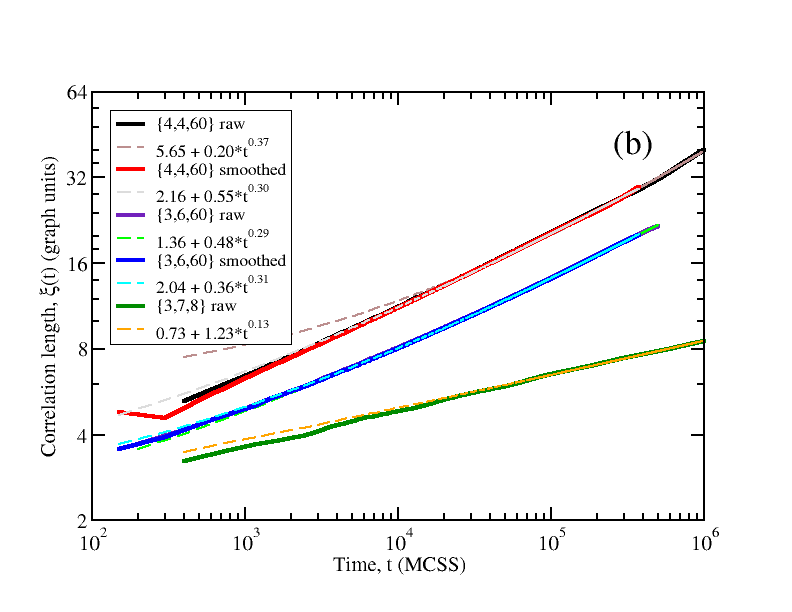}
\end{center}
\vspace{-1.0truecm}
\caption[]{
Simulation data for the correlation length $\xi$ (solid curves), together with 
three-parameter nonlinear fits (dashed curves), shown versus time $t$ 
for lattices $\{4,4,60\}$ (14,400 sites), $\{3,6,60\}$ (10,800 sites), and $\{3,7,8\}$ (11,173  sites).  
Linear scale (a) and log-log scale (b). 
In the legends, ``raw" refers to data not corrected for thermal fluctuations, 
and ``smoothed" refers to data with thermal fluctuations filtered out as 
discussed in the text. 
The filtered data give an estimate of the growth exponent of $n \approx 0.3$
for the Euclidean lattices, compatible with the theoretical value $n = 1/3$. 
However, the effective exponent obtained for the hyperbolic lattice is 
much lower: approximately 0.13. 
}
\label{fig-xi}
\end{figure}

The power-law result for the characteristic length scale given in 
Eq.~(\ref{eq-xi}) assumes an isotropic system with 
sharp interfaces and no thermal fluctuations 
in the bulk phase regions. Neither assumption is well satisfied for 
discrete Ising models at nonzero temperature. Care must therefore be 
exercised in extracting the relevant, growing length scale from the 
simulated spin configurations. 
Here we calculate the two-point correlation function, 
$G(r) = \langle s(r_i) s(r_i + r) \rangle$, where $r_i$ is the position of 
lattice point $i$, and $r_i+r$ is the position of a lattice point a distance 
$r$ away from $i$. 
Here, $r$ is defined as the shortest path between two lattice 
points along the edges (``taxicab" or ``Manhattan" distance). The correlation 
length $\xi(t)$ is estimated as the first zero crossing of $G(r)$ at time $t$.
See Fig.~\ref{fig-corrf}. 
To reduce the effect on the estimate of thermal fluctuations in the bulk phases,
we perform the simulations at relatively low temperatures, compared to $T_c$. 
In calculating the correlation functions we also ignore isolated single spins 
and spin pairs \cite{MAJU10}, which otherwise could distort $G(r)$ for $r=1$ 
and 2, as seen in Fig.~\ref{fig-corrf}.

\section{Numerical results}
\label{sec-Num}

\begin{figure}[h]
\vspace{-0.4truecm}
\begin{center}
\includegraphics[angle=0,width=.55\textwidth]{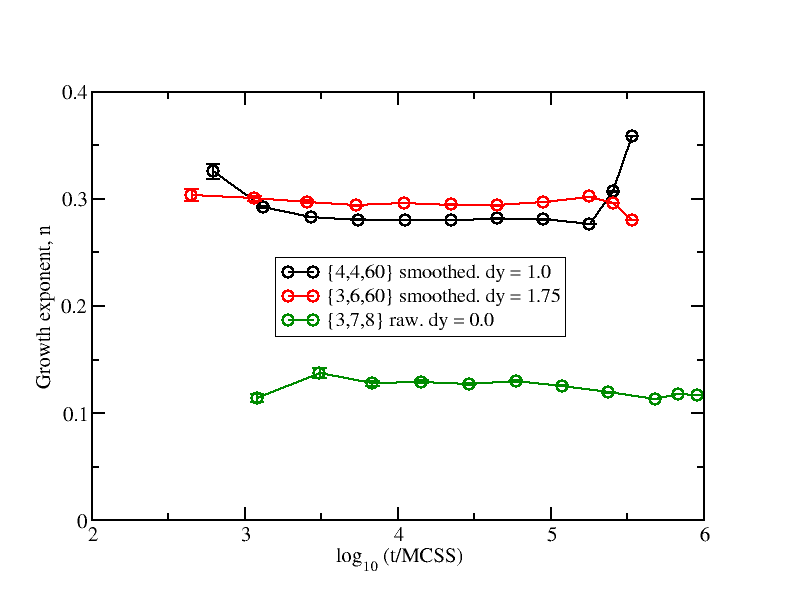}
\end{center}
\vspace{-1.0truecm}
\caption[]{
Estimates of the growth exponent $n$, based on the moving least-squares 
averaging method described in the text. The average values obtained are 
0.30 for the two Euclidean lattices, and 0.12 for the hyperbolic 
$\{3,7\}$ lattice. These values are consistent with the ones obtained 
by nonlinear fitting and shown in Fig.~\protect\ref{fig-xi}.  
$dy$ refers to the constant background subtraction applied to the data to 
obtain exponent estimates approximately independent of $t$.  
}
\label{fig-exp}
\end{figure}

The main numerical results of this study are summarized in 
Figs.~\ref{fig-xi} and \ref{fig-exp}. 
Figure \ref{fig-xi} shows the length scale $\xi$ as a function of time for each of the three studied lattices, 
following a quench to $T = 2.0$. The data sets were averaged over 
1,500 independent simulation runs for the Euclidean lattices, and 1000 runs
for the hyperbolic lattice. 
Results are included, both based on the raw data, and with thermal fluctuations filtered out as discussed above. 
Significant differences are only observed for $\{4,4\}$, for which the quench temperature is not very far below 
$T_c$. The growth exponents are here estimated by three-parameter, 
nonlinear fits to the time series, with the results 
given in the figure legends. For both the Euclidean lattices, 
the growth exponent comes out as $n \approx 0.3$, consistent 
with the expected value of $1/3$. 
For the hyperbolic $\{3,7\}$ lattice, however, the 
effective exponent is significantly lower, only about 0.13. 

In Fig.~\ref{fig-exp} we show results 
obtained by a different way of estimating the exponents. 
The data were divided into bins, each containing twice as many data points 
as the previous one (``octave binning"). We then performed a linear 
least-squares fit to $\log_{10} \xi$ versus $\log_{10} t$ 
in each successive pair of bins. Subtraction of a constant background was 
adjusted so that the estimated exponents became roughly independent of 
$t$. The resulting exponent 
estimates are seen to be consistent with those obtained by nonlinear 
fitting over the entire time interval. 

\begin{figure}[h]
\begin{center}
\includegraphics[angle=0,width=.5\textwidth]{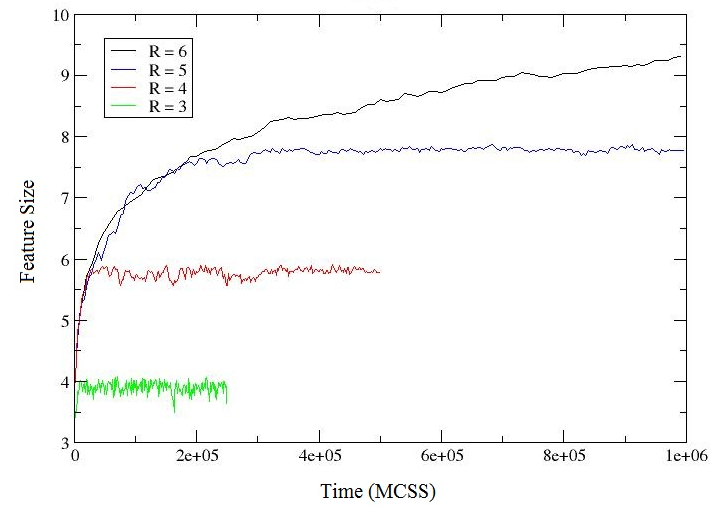}
\end{center}
\vspace{-0.7truecm}
\caption[]{
Correlation length $\xi$ versus time for $\{3,7\}$ lattices with 
$R = 3$,..., 6. These results indicate that the low value of the 
effective growth exponent observed with $R =8$ 
(see Figs.~\protect\ref{fig-xi} and \protect\ref{fig-exp}) 
is genuine and not due to finite-size saturation of the domain size 
within the simulation time of $10^6$ MCSS. 
}
\label{fig-scale}
\end{figure}

It is reasonable to ask whether the much lower 
effective growth exponent obtained for the hyperbolic lattice might 
be a result 
of finite-size saturation of the length scale. To check this possibility, 
we also performed simulations for smaller $\{3,7\}$ 
lattices with $R$ between 3 and 6. As seen in Fig.~\ref{fig-scale}, 
saturation does not appear to set in earlier than 
$10^6$ MCSS, even for a system as small as $R=6$. 

\section{Discussion}
\label{sec-Disc}

Here we have reported a preliminary, numerical investigation of 
the dynamics of phase separation in a model with a single, conserved, scalar 
order parameter (Model B) 
confined to the vertices of a lattice obtained as a 
regular tiling of a hyperbolic plane. The results are compared with those 
obtained for the same model on two different Euclidean lattices. The 
latter show power-law domain growth with an exponent of approximately 0.3,
near the theoretical result of 1/3, which is valid for the limiting case 
of an isotropic continuum system at zero temperature. Considering that the 
simulations were performed for anisotropic, discrete systems at a finite 
temperature, this agreement is convincing. 

However, for the hyperbolic $\{3,7\}$ lattice with $R$ up to 8, we observe 
much slower growth, consistent with a power law with an effective exponent 
of about 0.13. Whether or not this is indeed power-law growth or something 
else, we leave open for future, theoretical investigation. 
The effect is possibly related to the fact that for large $R$, most of the 
spins are located near the free surface. It may also be related to the 
mean-field nature of the phase transition at $T_c$ in the hyperbolic 
case, which indicates the existence of effective long-range interactions.

For the future 
we also leave a numerical investigation of phase ordering with non-conserved 
order parameter (Model A \cite{HOHE77}) on hyperbolic lattices. 
In this case, the growth exponent in the Euclidean case is known to be 1/2. 

\section*{Acknowledgments}
\label{sec-ack}

P.A.R.\ dedicates this paper to the memory of his beloved wife, Paulette Bond. 1949-2012. 

We thank Dr. Greg Brown for useful discussions and advice. 
This work was supported in part by 
U.S.\ National Science Foundation Grant Nos.\ OCI-1005117  
at Marshall University and DMR-1104829 at Florida State University.









\end{document}